\documentclass[sigconf]{acmart}
\usepackage[ruled,linesnumbered]{algorithm2e}
\usepackage{graphicx}
\usepackage{subfig}
\usepackage{threeparttable}
\usepackage{booktabs}
\usepackage{multirow}
\usepackage{caption}
\copyrightyear{2020} 
\acmYear{2020} 
\setcopyright{acmcopyright}
\acmConference[KDD '20]{Proceedings of the 26th ACM SIGKDD Conference on Knowledge Discovery and Data Mining}{August 23--27, 2020}{Virtual Event, CA, USA}
\acmBooktitle{Proceedings of the 26th ACM SIGKDD Conference on Knowledge Discovery and Data Mining (KDD '20), August 23--27, 2020, Virtual Event, CA, USA}
\acmPrice{15.00}
\acmDOI{10.1145/3394486.3403201}
\acmISBN{978-1-4503-7998-4/20/08}

\begin{document}
\title{Multi-level Graph Convolutional Networks for Cross-platform Anchor Link Prediction}

\author{Hongxu Chen}
\affiliation{%
  \institution{University of Technology Sydney}
}
\email{hongxu.chen@uts.edu.au	}

\author{Hongzhi Yin}
\authornote{Corresponding author and having equal contribution with the first author.}
\affiliation{%
  \institution{The University of Queensland}
}
\email{h.yin1@uq.edu.au}

\author{Xiangguo Sun}
\affiliation{%
  \institution{Southeast University}
}
\email{sunxiangguo@seu.edu.cn}

\author{Tong Chen}
\affiliation{%
  \institution{The University of Queensland}
}
\email{tong.chen@uq.edu.au}

\author{Bogdan Gabrys}
\affiliation{%
  \institution{University of Technology Sydney}
}
\email{Bogdan.Gabrys@uts.edu.au}

\author{Katarzyna Musial}
\affiliation{%
  \institution{University of Technology Sydney}
}

\email{Katarzyna.Musial-Gabrys@uts.edu.au}

\begingroup
\mathchardef\UrlBreakPenalty=10000

\renewcommand{\shortauthors}{H. Chen, H. Yin, et al.}

\begin{abstract}
Cross-platform account matching plays a significant role in social network analytics, and is beneficial for a wide range of applications. However, existing methods either heavily rely on high-quality user generated content (including user profiles) or suffer from data insufficiency problem if only focusing on network topology, which brings researchers into an insoluble dilemma of model selection. In this paper, to address this problem, we propose a novel framework that considers multi-level graph convolutions on both local network structure and hypergraph structure in a unified manner. The proposed method overcomes data insufficiency problem of existing work and does not necessarily rely on user demographic information. Moreover, to adapt the proposed method to be capable of handling large-scale social networks, we propose a two-phase space reconciliation mechanism to align the embedding spaces in both network partitioning based parallel training and account matching across different social networks. Extensive experiments have been conducted on two large-scale real-life social networks. The experimental results demonstrate that the proposed method outperforms the state-of-the-art models with a big margin.

\end{abstract}


\maketitle

\section{Introduction}
Nowadays, most people participate in more than one Online Social Network (OSN), such as Facebook, Twitter, Weibo, Linkedin. More often than not, users sign up at different OSNs for different purposes, and different OSNs show different views and aspects of people. For example, a user makes connections to their friends on Facebook, but uses Linkedin to connect to his/her colleagues, interested companies and seek job opportunities. Though different OSNs exhibit distinct features and functionalities, a large portion of overlapping individual user accounts across different social platforms have been always witnessed. However, the information about multiple accounts that belong to the same individual is not explicitly given in most social networks due to either privacy concerns or lack of motivation \cite{man2016predict, musial2013social}.

The problem of matching accounts that belong to the same individual from different social networks is defined as \textbf{Account Mapping} \cite{tan2014mapping}, \textbf{Social Network De-anonymization}\cite{zhou2015cross, zhang2015cosnet, narayanan2009anonymizing} or \textbf{Social Anchor Link Prediction} \cite{zhang2015integrated,man2016predict,cheng2019deep} in Data Mining research field. Account Matching across different social platforms plays a fundamental and significant role in social network analytics as it helps improve many downstream applications, such as online personalized services \cite{cao2016bass}, link prediction \cite{ahmad2010link}, recommender systems \cite{man2015context, tang2012etrust, yin2018joint,yin2019social}, biology protein-protein alignment for ageing related complexes \cite{faisal2014global}, and criminal behaviour detection \cite{tan2014mapping}. Although much attention has been dedicated to this challenging subject, there is still plenty of room for improvement. Previous studies \cite{riederer2016linking, liu2013s, iofciu2011identifying, malhotra2012studying} proposed to solve this problem by exploiting available auxiliary information such as self-generated user profiles, daily generated content and other demographic features (e.g., user name, profile picture, location, gender, post, blogs, reviews, etc.). However, with the increased public awareness of privacy and information rights, these information is becoming less available and accessible.  

Recently, with the advances in Network Embedding (NE) techniques, research attention related to this problem has been shifted to focus on mining network structure information \cite{tan2014mapping, man2016predict, liu2016aligning, cheng2019deep} as it has been claimed that the social network structural data is much more reliable in terms of correctness and completeness. However, only focusing on modelling the network structure itself makes almost all existing methods suffer from data insufficiency problems, especially in small-scale networks and cold-start settings (i.e., a user is new to the network). Therefore, it has been a dilemma confronting practitioners in the real-world scenarios, and effective solutions are urgently needed. 

In light of this, we propose to exploit and integrate the hypergraph information distilled from the original network for data enhancement. In the rest of the paper, we use the terms ``simple graph'' and ``hypergraph'' to denote original network and hypergraphs extracted from original network, respectively. Compared to simple graphs,  hypergraphs allow one edge (a.k.a., heperedge) to connect more than two nodes simultaneously. This means non-pairwise relations among nodes in a graph can be easily organized and represented as hyperedges. Moreover, hypergraphs are robust, flexible and can fit a wide variety of social networks, no matter the given networks are pure social networks or heterogenous social networks with various types of attributes and links.

More specifically, we propose a novel embedding framework \textbf{Multi-level Graph Convolutional Networks}, namely \textbf{MGCN}, to jointly learn embeddings for network vertices at different levels of granularity w.r.t. flexible GCN kernels (i.e., simple graph GCN, hypergraph GCN). Simple graph structure information of social networks reveals relationships among users (e.g., friendships, followers), while hypergraphs carry different semantic meanings depending on their specific definitions in a social network. For example, N-hop neighbour-based hypergraphs (N-hop neighbours of a user are connected via a same hyperedge) represent friends circle in some extent. Centrality-based hypergraphs represent different social levels (users with similar centrality values may be of same social status). Therefore, by defining various hypergraphs and intergating them into network embedding learning will facilitate learning better user representations. To support this, our proposed MGCN framework is flexible and can incorporate various hypergraph definitions, which can take any hypergraphs as vector representations, making the model structure invariant to various hypergraph definitions.

 The rationale behind exploiting and integrating hypergraphs by extending GCN is that hypergraphs provide a more flexible network representation that can contain additional and richer information compared to individual, single graph GCNs on local network topology. It has been found that the optimal number of GCN layers is always set to two in most cases because adding more layers cannot significantly improve the performance \cite{hamilton2017inductive}. As a result, GCNs are only able to capture the local information around a node in networks. This phenomenon also makes solo GCN contradictory and thus perform mediocrely on account matching task as the key to the task is to explore more and deeper information to make the predictions. Intuitively, defining GCNs on hypergrpahs extracted from original networks will be complementary to the limitations of existing GCN-based network embedding models. 
 
 Nevertheless, it is still a challenging task because social networks are large-scale with millions of nodes and billions of edges. Traditional centralized training methods fail to scale for such large networks, due to high computation demands. To adapt MGCN for large scale social networks, and improve its scalability and efficiency, we propose a novel training method that first partitions the large-scale social networks into clusters and learns network embeddings in a fully decentralized way. To align the learned embedding spaces of different clusters, we propose a novel two-phase space reconciliation mechanism. At the first stage, we align the embedding spaces learned from each cluster within the same network. In addition to the alignment between different subnetworks in the same network, the second-phase space reconciliation aligns two different networks through a small number of observed anchor nodes, which makes our MGCN framework achieve more accurate anchor link prediction than state-of-the-art models and high efficiency on large social networks.

The main contributions of this paper are summrized as follows:
\begin{itemize}
	\item We propose a novel framework for the challenging task of predicting anchor links across different social networks. The proposed method MGCN takes both local and hypergraph level graph convolutions into consideration to learn network embeddings, which is able to capture wider and richer network information for the task.
	\item In order to adapt the proposed framework to be able to cope with large scale social networks, we propose a series of treatments including network partitioning and space reconciliation to handle the distributed training process. 
	\item Extensive evaluations on large-scale real-world datasets have been conducted, and the experimental results demonstrate the superiority of the proposed MGCN model against state-of-the-art models. 
\end{itemize}
\vspace{-1em}


\section{Proposed Method}
\vspace{-.5em}
\subsection{Preliminaries} 
\subsubsection{\textbf{Problem Definition}} 
Given a pair of networks $\mathcal{G}_1=\{\mathcal{V}_1,\mathcal{E}_1\}$ and $\mathcal{G}_2=\{\mathcal{V}_2,\mathcal{E}_2\}$, and a set of observed anchor links $\mathcal{S}_{anchor}=\{(u,v)|u\in \mathcal{V}_1, v\in \mathcal{V}_2\}$, our goal is to predict those unobserved anchor links across $\mathcal{G}_1$ and $\mathcal{G}_2$. We treat this task as binary classification, that is, given a pair of nodes $(u,v)$ where $u\in \mathcal{V}_1, v\in \mathcal{V}_2$, we predict if there is a link between them.

\subsubsection{\textbf{Hypergraph}} 
\par In simple graphs, an edge connects two nodes, while an edge in a hypergraph (i.e. hyperedge) can connect more than two nodes. We denote a hypergraph by $\mathcal{G}^h=\{\mathcal{V},\mathcal{E}^h\}$, where $\mathcal{V}$ is the node set, $\mathcal{E}^h$ is the hyperedge set. For each hyperedge $e \in \mathcal{E}^h$, we have $e=\{v_1,\cdots,v_p\}, v_i \in \mathcal{V},2<p\leq |\mathcal{V}|$.
\vspace{-1.1em}
\subsection{Model Overview}
To predict anchor links, we introduce a novel multi-level graph convolutional network (MGCN) to learn the embeddings of each network. Figure \ref{fig:Muli_level_conv} is an illustration of our proposed MGCN framework, which consists of two levels of graph convolution operations. It firstly performs convolution on simple graphs (i.e., the original social network in our case). After obtaining the node embeddings from the simple graph convolution,  the node embeddings are refined by an innovative convolution operation defined on hypergraphs. With the final embeddings of two social networks obtained, we align the latent space of two networks via an embedding reconciliation process. Lastly, we deploy a fully connected network to predict whether an anchor link exists between any arbitrary pair of nodes from two networks. 
In addition, we present a parallelizable scheme that allows MGCN to efficiently handle large-scale networks through graph partitioning.
\begin{figure}[t] 
\centering
\includegraphics[width=0.5\textwidth]{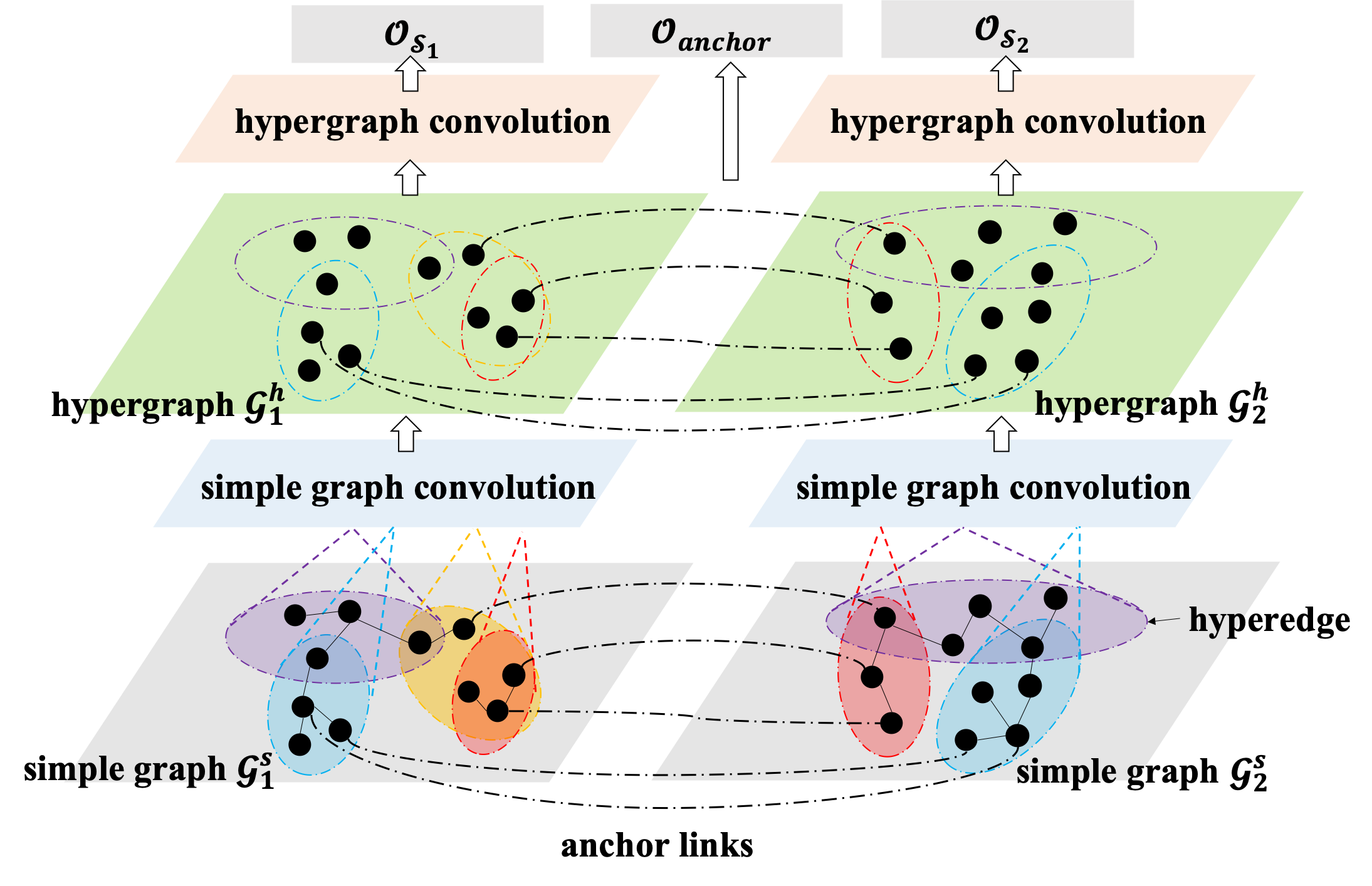}
\vspace{-3em}
\caption{Multi-level graph Convolution.}
\label{fig:Muli_level_conv}
\vspace{-2em}
\end{figure}
\vspace{-1em}
\subsection{Convolution on Simple Graphs}
Given an original social network $\mathcal{G}=\{\mathcal{V},\mathcal{E}\}$ (i.e., simple graph), assume that we have constructed a hypergraph $\mathcal{G}^h$ from $\mathcal{G}$, where each hyperedge $e\in \mathcal{E}^h, e=\{v_1,v_2,\cdots,v_n\},v_i\in \mathcal{V}$. 
We first perform simple graph convolutions in order to obtain the base embeddings of all nodes, denoted by $\mathbf{X}\in \mathbb{R}^{|\mathcal{V}\times d|}$, where $d$ is the dimension of each node embedding vector. We start with a simple graph convolution within hyperedge $e$ by:
\begin{equation}\label{eq:simple_GCN}
	\mathbf{X}^{k+1}_e=\sigma(\mathbf{A}_e\mathbf{X}_e^{k}\mathbf{W}^{k})
\end{equation}
where $\mathbf{A}_e \in \mathbb{R}^{|\mathcal{V}|\times |\mathcal{V}|}$ is the adjacency matrix within hyperedge, $\sigma(\cdot)$ denotes the non-linear activation function such as $ReLU(\cdot) = \max(0,\cdot)$, while $\mathbf{X}^k_{e}$ and $\mathbf{W}^k$ carry the latent representations and the trainable weights in the $k$-th convolution layer. Specifically, in contrast to the plain GCN \cite{kipf2016semi} that simply operates on the entire graph, we perform the convolution operation on each hyperedge individually. The rationale is that we can incorporate fine-grained local structural information from the hyperedges into the learned node embeddings. To achieve this, we define a diagonal matrix $\mathbf{S}_e\in \mathbb{R}^{|\mathcal{V}|\times |\mathcal{V}|}$ for hyperedge $e$, where each entry $\mathbf{S}_e(v_i, v_j)$ is:
\begin{equation}
\begin{split}
\mathbf{S}_e(v_i,v_j)=\left\{
             \begin{array}{ll}
             p(v,e),& \text{if } v_i=v_j, v_i \in e \\
             0, &\text{otherwise} 
             \end{array}
\right.
\end{split}
\end{equation}
where $p(v,e)$ stands for the possibility of observing node $v$ in hyperedge $e$, and its calculation depends on particular definitions of hyperedges (see Section \ref{subsec:hyperconst} for possible options). Then, let $\hat{\mathbf{A}}=\mathbf{I}_{|\mathcal{V}|}+\mathbf{D}^{-\frac{1}{2}}\mathbf{A}\mathbf{D}^{-\frac{1}{2}}$ where $\mathbf{D}\in \mathbb{R}^{|\mathcal{V}|\times |\mathcal{V}|}$ is a diagonal matrix containing each node's degree in simple graph $\mathcal{G}$, $\mathbf{A}$ is the adjacency matrix of the simple graph $\mathcal{G}$, and $\mathbf{I}_{|\mathcal{V}|}$ is the identity matrix. Then, the local adjacency matrix $\mathbf{A}_e$ for hyperedge $e$ is calculated via: 
\begin{equation}
	\mathbf{A}_e=\mathbf{S}_e\hat{\mathbf{A}}\mathbf{S}_e
\end{equation}

Intuitively, $\mathbf{A}_e$ can be viewed as an adjacency matrix for the directly connected nodes in hyperedge $e$, which is further weighted by the hyperedge connectivity in $\mathbf{S}_e(v_i, v_j)$. As a result, when performing simple graph convolutions, we can simultaneously take two types of local node-node structural information into consideration, making the learned base embeddings more expressive. Based on Equation~\ref{eq:simple_GCN}, the convolution operation on the entire simple graph $\mathcal{G}$ can be obtained through the summation across all hyperedges:
\begin{equation}
	\mathbf{X}^{k+1}_{simple}=f( \oplus_{e\in \mathcal{E}^h}\mathbf{X}^{k+1}_e)
\end{equation}
where $\oplus$ means the concatenation of the output for each hyperedge $e$, and $f(\cdot)$ denotes a dense layer that maps the concatenated embeddings back to a $d$-dimensional space. 
\vspace{-0.5em}
\subsection{Convolution on Hypergraphs} \label{sub:hyper}
With the base embeddings $\mathbf{X}^{K}_{simple}$ learned in the simple graph convolution stage for the final $K$-th convolution layer, we further infuse the structural information of the constructed hypergraph $\mathcal{G}^h$ into every node's latent representation. In recent years, hypergraph convolution network has started to attract attention from the network embedding research community \cite{feng2019hypergraph,yadati2019hypergcn,jiang2019dynamic}. Different from most related works that deduce hypergraph convolution using the spectral convolution theory, we derive the mathematical form of hypergraph convolution by treating it as a generalized version of simple graph convolution, which makes the inference process more intuitive and natural to understand.

\par Given a hypergraph $\mathcal{G}^h=\{\mathcal{V},\mathcal{E}^h \}$, let $\mathbf{H}\in \mathbb{R}^{|\mathcal{V}|\times |\mathcal{E}^h|}$ be an incidence matrix where each entry $\mathbf{H}(v,e)$ is determined by:
\begin{equation}
\begin{split}
\mathbf{H}(v,e)=\left\{
             \begin{array}{ll}
             p(v,e),& \text{if } v \in e \\
             0, &\text{otherwise} 
             \end{array}
\right.
\end{split}
\end{equation}
where $p(v,e)$ indicates the possibility that node $v$ belongs to hyperedge $e$. Let the diagonal matrix $\mathbf{D}_n\in \mathbb{R}^{|\mathcal{V}|\times |\mathcal{V}|}$ denoting the degree of nodes in the hypergraph such that $\mathbf{D}_n(v,v)=\sum_{e\in \mathcal{E}^h}\mathbf{H}(v,e)$. Similarly, the degree of hyperedges can be denoted by a diagonal matrix $\mathbf{D}_e\in \mathbb{R}^{|\mathcal{E}^h|\times |\mathcal{E}^h|}$ where $\mathbf{D}_e(e,e)=\sum_{v\in \mathcal{V}}\mathbf{H}(v,e)$. Since $\mathbf{H}$ indicates the correlation between nodes and hyperedges, we can use $\mathbf{H}\mathbf{H}^{\top}$ to quantify the pairwise relationships between nodes. Then, the weighted adjacency matrix $\mathbf{A}_h\in\mathbb{R}^{|\mathcal{V}|\times|\mathcal{V}|}$ of hypergraph $\mathcal{G}^h$ can be derived as:
\begin{equation}\label{equ:adj_hyper}
	\mathbf{A}_h=\mathbf{H}\mathbf{H}^{\top}-\mathbf{D}_n
\end{equation} 

Having acquired the adjacency matrix of hypergraph, we can naturally extend simple graph convolution to hypergraph $\mathcal{G}^h$. Recall that in the typical GCN framework presented in \cite{kipf2016semi}, for a simple graph $\mathcal{G}^s=\{\mathcal{V}^s,\mathcal{E}^s\}$, the standard graph convolution is defined as:
\begin{equation}\label{eq:plain_GCN}
	\mathbf{X}^{k+1}_s=\sigma \left( \Big{(} \mathbf{I}_{|\mathcal{V}^s|}+\mathbf{D}_s^{-\frac{1}{2}}\mathbf{A}_s\mathbf{D}_s^{-\frac{1}{2}} \Big{)}\mathbf{X}_s^{k}\mathbf{W}^k_s \right)
\end{equation} 
where $\mathbf{D}_s$ contains all nodes' degree of $\mathcal{G}^s$, $\mathbf{A}_s$ is the adjacency matrix of $\mathcal{G}^s$. Apart from the identity matrix $\mathbf{I}_{|\mathcal{V}^s|}$, the above standard graph convolution, at its core, are dependent on the node relationships encoded in the degree and adjacency matrices $\mathbf{D}_s$ and $\mathbf{A}_s$. Therefore, by replacing its input with the corresponding information extracted from the hypergraph $\mathcal{G}^h$, we can effectively model hypergraph convolution in a similar way to the standard GCN at each layer $k$:
\begin{equation}
	\begin{aligned}
		\mathbf{X}^{k+1}=&\sigma \left( \Big{(}\mathbf{I}_{|\mathcal{V}|}+\mathbf{D}_n^{-\frac{1}{2}}\mathbf{A}^h\mathbf{D}_n^{-\frac{1}{2}} \Big{)} \mathbf{X}^{k}\mathbf{W}^k\right)\\
		=&\sigma \left( \Big{(} \mathbf{I}_{|\mathcal{V}|}+\mathbf{D}_n^{-\frac{1}{2}}\left(\mathbf{H}\mathbf{H}^{\top}-\mathbf{D}_n\right)\mathbf{D}_n^{-\frac{1}{2}} \Big{)} \mathbf{X}^{k}\mathbf{W}^k \right)\\
		=&\sigma \Big{(}\mathbf{D}_n^{-\frac{1}{2}}\mathbf{H}\mathbf{H}^{\top}\mathbf{D}_n^{-\frac{1}{2}}\mathbf{X}^{k}\mathbf{W}^k \Big{)}
	\end{aligned}
\end{equation}
Let $\mathbf{\Theta} = \mathbf{D}_v^{-\frac{1}{2}}\mathbf{H}\mathbf{H}^T\mathbf{D}_v^{-\frac{1}{2}}$, then we have:
\begin{equation}\label{eq:final}
	\begin{aligned}
		\mathbf{X}^{k+1}=\sigma (\mathbf{\Theta} \mathbf{X}^{k}\mathbf{W}^k)
	\end{aligned}
\end{equation}
where $\mathbf{X}^{k} = \mathbf{X}^{K}_{simple}$ when $k=0$. Suppose we also adopt $K$ layers of convolution on hypergraph, then the final output of the multi-level graph convolutional network is denoted by $\mathbf{X}^K$. By this mean, the generated node embeddings in $\mathbf{X}^{k+1}_{simple}$ can both capture pairwise relations (i.e., 1-hop neighbourhood) and high-order non-pairwise relations (i.e., hyperedges). As we will further discuss in Section \ref{subsub:observed_per}, this is especially important when the number of observed anchor nodes for training are limited.

\begin{figure*}[t]
\vspace{-1em}
\centering
\includegraphics[width=0.8\textwidth]{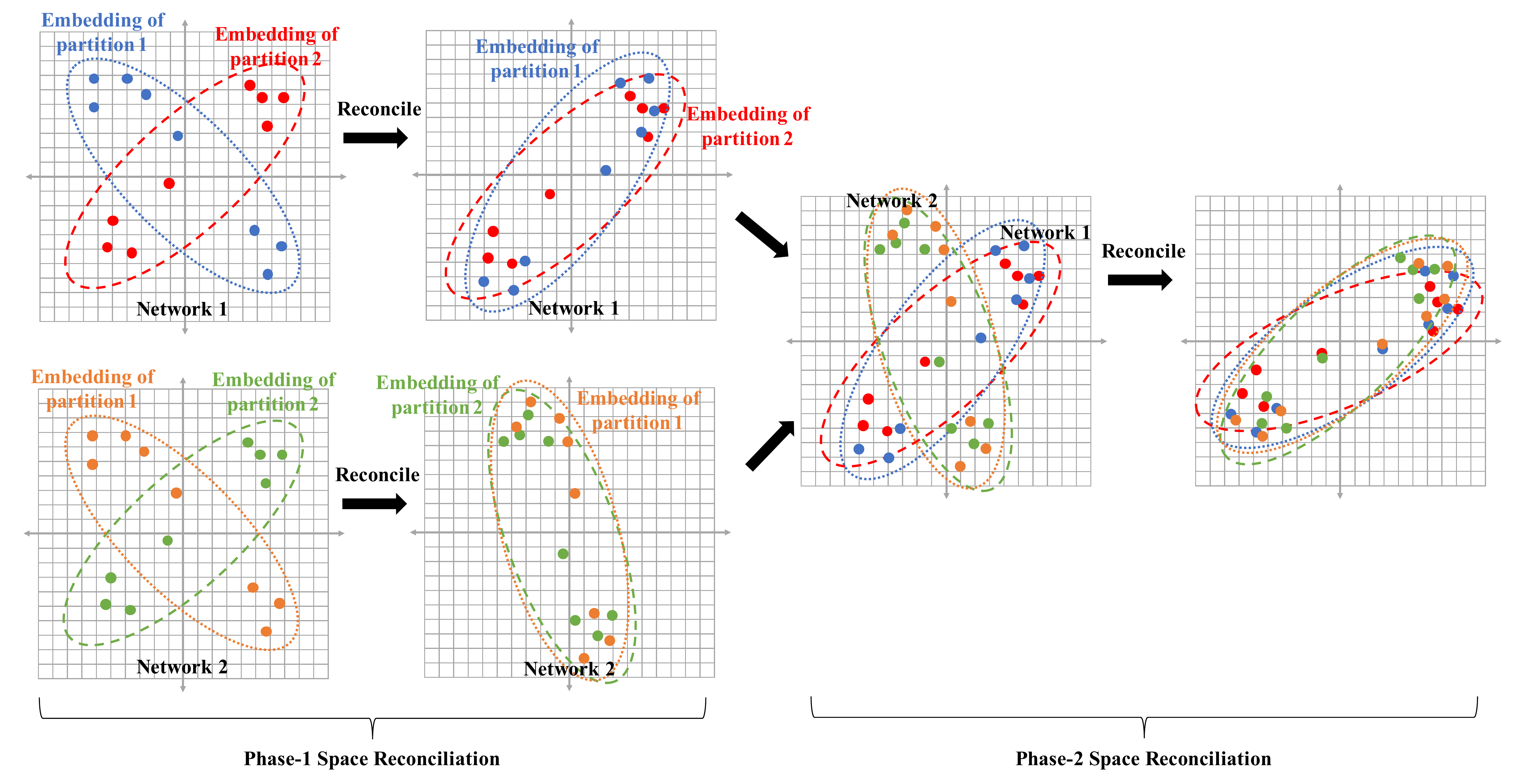}
\vspace{-1em}
\caption{Two-phase embedding space reconciliation.}
\label{fig:reconcile}
\vspace{-1.5em}
\end{figure*}

\vspace{-1em}
\subsection{Learning Network Embeddings}
For network embedding, the output embeddings from Equation \ref{eq:final} are learned by maximizing the probability of positive edges and minimizing the probability of negative ones:
\begin{equation}\label{equ:emb_loss}
\begin{aligned}
\mathcal{O}_{embedding} &=\sum_{(v_i, v_j) \in \mathcal{E}} \log \eta (\mathbf{x}_{i}^{K\top} \mathbf{x}_{j}^{K}).\\
&+\sum_{k=1}^{M} E_{v_{k} \propto P(v)}\Big{[}\log \big{(}1-\sigma(\mathbf{x}_{i}^{K\top}\mathbf{x}_{k}^{K})\big{)}\Big{]} \\
&+\sum_{k=1}^{M} E_{v_{k} \propto P(v)}\Big{[}\log \big{(}1-\sigma(\mathbf{x}_{j}^{K\top} \mathbf{x}_{k}^{K})\big{)}\Big{]}
\end{aligned}
\end{equation}

where $\eta(\cdot, \cdot)$ is the sigmoid function to calculate the probability of observing edge $(v_i,v_j)$.

For a given positive edge $(v_i,v_j)$ in the training set, we use bidirectional negative sampling strategy \cite{chen2018pme} to draw negative edges for training. Specifically, we fix $v_i$ and generate $M$ negative nodes $v_k$ via a noise distribution $P_n(v)\sim d_v^{0.75}$, where $d_v$ is the degree of node $v$. Then we fix $v_j$ and sample $M$ negative nodes with the same process. By optimizing Equation \ref{equ:emb_loss}, we can obtain optimal embeddings in $\mathbf{X}^K$ from the last layer $K$. Afterwards, the final embeddings are further leveraged for downstream anchor link prediction task.
\vspace{-1em}
\begin{algorithm}[]
\KwIn{
 $\mathcal{G}(\mathcal{V},\mathcal{E})$, $N_{max}$, $N_{min}$, iteration $T$.
}
\KwOut{
partitions $P=\{\mathcal{G}_1(\mathcal{V}_1,\mathcal{E}_1),\cdots,\mathcal{G}_n(\mathcal{V}_n,\mathcal{E}_n)\}$.
}
$P=\text{Louvain}(\mathcal{G})$ //Generating partitions $P$ from $\mathcal{G}$ according to Louvain algorithm\cite{blondel2008fast}. \\
\For{iter from $1$ to $T$}
{
\For{partition $\mathcal{G}^{'} \in P$}
{
\uIf{$|\mathcal{V}^{'}|<N_{min}$}
{add nodes of $\mathcal{V}^{'}$ into other partitions, delete $\mathcal{G}^{'}$.}
\uElseIf{$N_{min}<|\mathcal{V}^{'}|<=N_{max}$}
{continue}
\Else{
$P_{t}=\text{Louvain}(\mathcal{G}^{'})$ //Generating partitions $P_t$ from $\mathcal{G}^{'}$ according to Louvain algorithm \cite{blondel2008fast}.\\
$P=P \cup P_t$
}
}
}
\Return{$P$}
\caption{Graph Partitioning}
\label{alg:partition}
\vspace{-1ex}
\end{algorithm}
\vspace{-2.4em}
\subsection{Anchor Link Prediction} \label{sub:anchor_link}
Note that after acquiring the final representations $\mathbf{X}_1^K$ and $\mathbf{X}_2^K$ of two networks $\mathcal{G}_1$ and $\mathcal{G}_2$, we should not directly use them for anchor link prediction because the node representations are learned in two different latent spaces, which may vary a lot in terms of semantic contexts. Instead, we first reconcile both of them into the same latent space, and then use the aligned embeddings for anchor link prediction. To reconcile $\mathbf{X}_1^K$ and $\mathbf{X}_2^K$ into the same space, we fix $\mathbf{X}_1^K$ and project $\mathbf{X}_2^K$ into the same space as $\mathbf{X}_1^K$. Let $\gamma (.|\mathbf{\Gamma},\mathbf{b} ) $ be a projection function with a projection matrix $\mathbf{\Gamma}$ and bias $\mathbf{b}$. Then, by aligning the embedding vectors of the anchor nodes in both graphs, we can learn the parameters in the projection function, thus ensuring accurate reconciliation for two latent spaces:
\begin{equation}\label{equ:matching}
\mathcal{O}_{anchor}=\!\!\!\!\!\!\!\!\sum_{(v, u) \in \mathcal{S}_{anchor}}\!\!\!\!\!\!\!\!\!\!\! \Arrowvert \mathbf{X}_1^{K}[v,:]-\phi (\mathbf{X}_2^{K}[u,:]\arrowvert\mathbf{\Theta},\mathbf{b} )\Arrowvert^2
\end{equation}
where $\gamma (\mathbf{x}|\mathbf{\Gamma},\mathbf{b})=\mathbf{x}\mathbf{\Gamma}+\mathbf{b} $, and $\mathcal{S}_{anchor}$ is the labeled anchor links. Then, for any pair of nodes $(v_i,v_j), v_i\in \mathcal{G}_1, v_j \in \mathcal{G}_2$, the representation of this pair can be denoted by the concatenation of their corresponding embeddings. We sent these pair embeddings into a fully connected network and finally output the prediction of whether they are anchor link, and use cross entropy as the loss function of anchor link prediction. 
\vspace{-0.5em}
\subsection{Handling Large-Scale Networks}\label{sub:large}
\par Although GCN-based methods  have been widely used in various tasks, most related methods still suffer from the ``last mile'' technology when we deal with large-scale networks because most GCN-based methods need the global adjacency matrix as their inputs, and this easily causes out of memory issues for GPU computation. Besides, when the network scale increases, it will also lead to growth in computation time. Thus, we need an effective graph partition strategy so that we can deploy the proposed MGCN in parallel. To this end, we first present a graph partitioning approach via Algorithm \ref{alg:partition}, and propose a two-phase reconciliation mechanism as shown in Figure \ref{fig:reconcile}. Specifically, we split the large network into several partitions according to modularity maximization, and then deploy our model in every single partition. For each graph, we reconcile the latent spaces of all its partitions into the same one using the reserved anchor nodes when partitioning the whole graph. Then, we align the embeddings of $\mathcal{G}^1$ and $\mathcal{G}^2$ into the same latent space using observed anchor nodes from two graphs.

\subsubsection{\textbf{Graph Partition}} As Algorithm \ref{alg:partition} depicts, to split the large network into several partitions with acceptable size (from $N_{min}$ nodes to $N_{max}$ nodes, for example), we first compute the partition of the network which maximises the modularity using the Louvain algorithm\cite{blondel2008fast}. For each partition $\mathcal{G}^{'}=\{\mathcal{V}^{'},\mathcal{E}^{'}\}$, if the size is larger than the upper bound, that is $|\mathcal{V}^{'}|>N_{max}$, we put $\mathcal{G}^{'}$ as the input again and repeat the algorithm to further split $\mathcal{G}^{'}$ into more smaller partitions. If $|\mathcal{V}^{'}|<N_{min}$, we randomly assign it to other created partitions.
\vspace{-.5em}
\subsubsection{\textbf{Reconcile Latent Embedding Spaces}}
We have noticed that to deploy our model into different partitions independently actually produce the embeddings in different latent spaces. Therefore we need to further match different partitions into the same representation space. Here, we select $N$ nodes from the network as shared nodes across all partitions, and append these $N$ nodes as well as their associated edges into all partitions. Then we select one of the partitions as a fixed one, and reconcile the others into the same space with it. For example, for all $P$ partitions $\{\mathcal{G}_1,\mathcal{G}_2,\cdots,\mathcal{G}_P\}$, we fix partition $\mathcal{G}_1$ and all other partitions' embeddings are transformed via a linear function $g(.)$. We maximize the following target:
\vspace{-.5em}
\begin{equation}
\begin{aligned} 
\mathcal{O}_{partition}&=\sum_{p=2}^P\sum_{v_i\in \mathcal{V}_{shared}}\log \sigma \left((f_p(\mathbf{x}^{(p)}_i))^\top \mathbf{x}^{(1)}_i \right)
\end{aligned}\vspace{-.5em}
\end{equation}
where $ \mathcal{V}_{shared}$ is the set of shared nodes appearing in all partitions, and $\mathbf{x}^{(p)}_i$ is the representation of node $v_i$ in partition $p$. Having matched each partition into the same space, we can get the final network embeddings in a uniform space, we can eventually use them as described in Section \ref{sub:anchor_link} to predict the anchor links.
\vspace{-.8em}
\subsection{Optimization Strategy}
We train MGCN model in a step-by-step manner. Specifically, we first train MGCN by optimizing the graph embedding objective function $\mathcal{O}_{embedding}$. After that, we optimize the graph partition reconciliation objective function $\mathcal{O}_{partition}$ (i.e., phase-1 space reconciliation), then optimize the reconciliation objective $\mathcal{O}_{anchor}$ (i.e., phase-2 space reconciliation). Lastly, with the fully aligned node embeddings from both graphs, we optimize MGCN for the anchor link prediction task by minimizing the cross-entropy loss.

\vspace{-1em}
\section{Experiments}

\begin{figure*}[h!]
\vspace{-2.5em}
\centering
\subfloat[Anchor Link Prediction on Facebook-Twitter]{
\label{fig:anchor}
\hspace{-0.5cm}
\includegraphics[width=5.7cm, height=3cm]{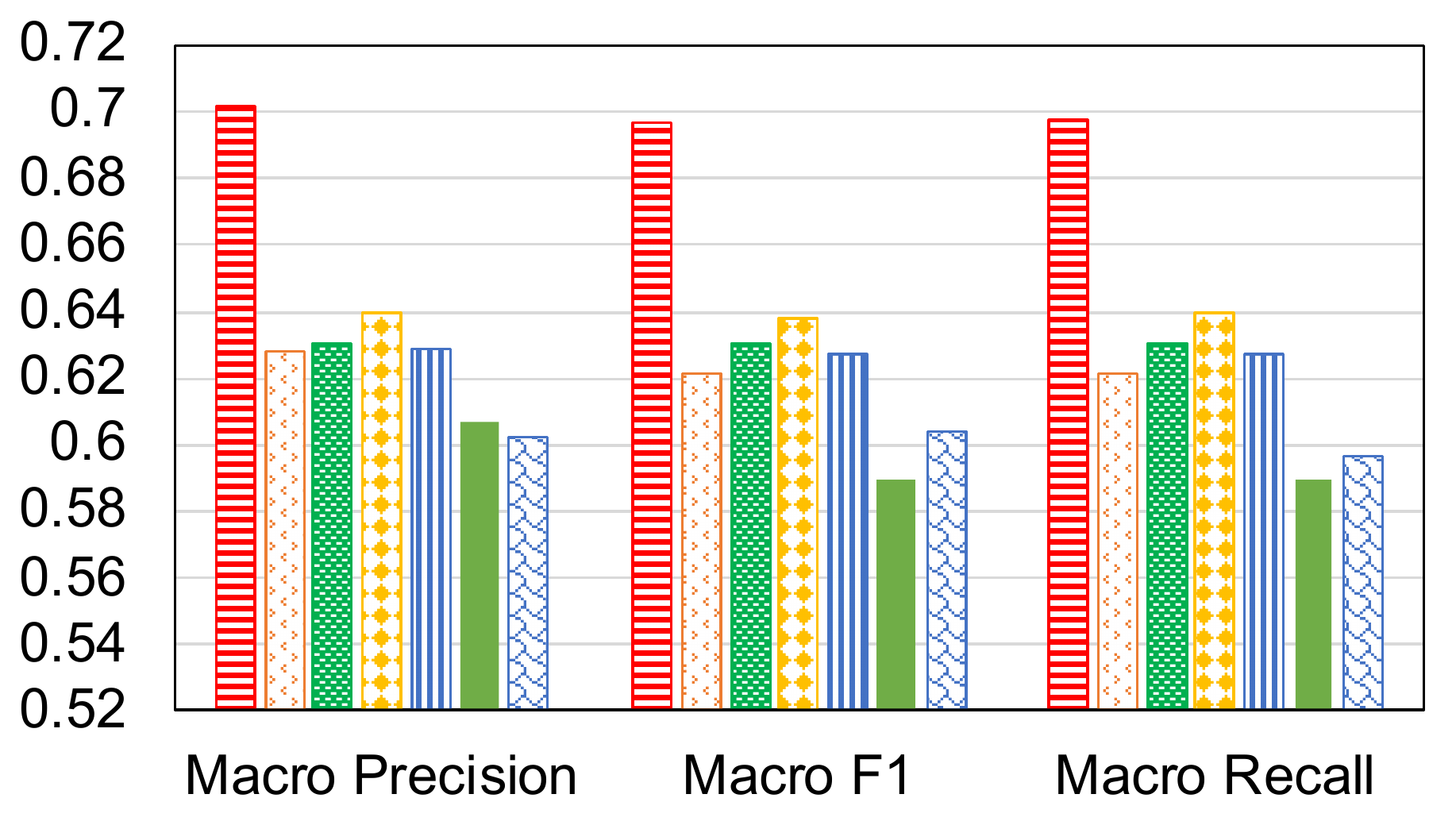}%
}
\hspace{-2.0cm}
\subfloat[Anchor Link Prediction on Douban-Weibo]{
\label{fig:link}
\hspace{2.0cm}
\includegraphics[width=8cm, height=3.09cm]{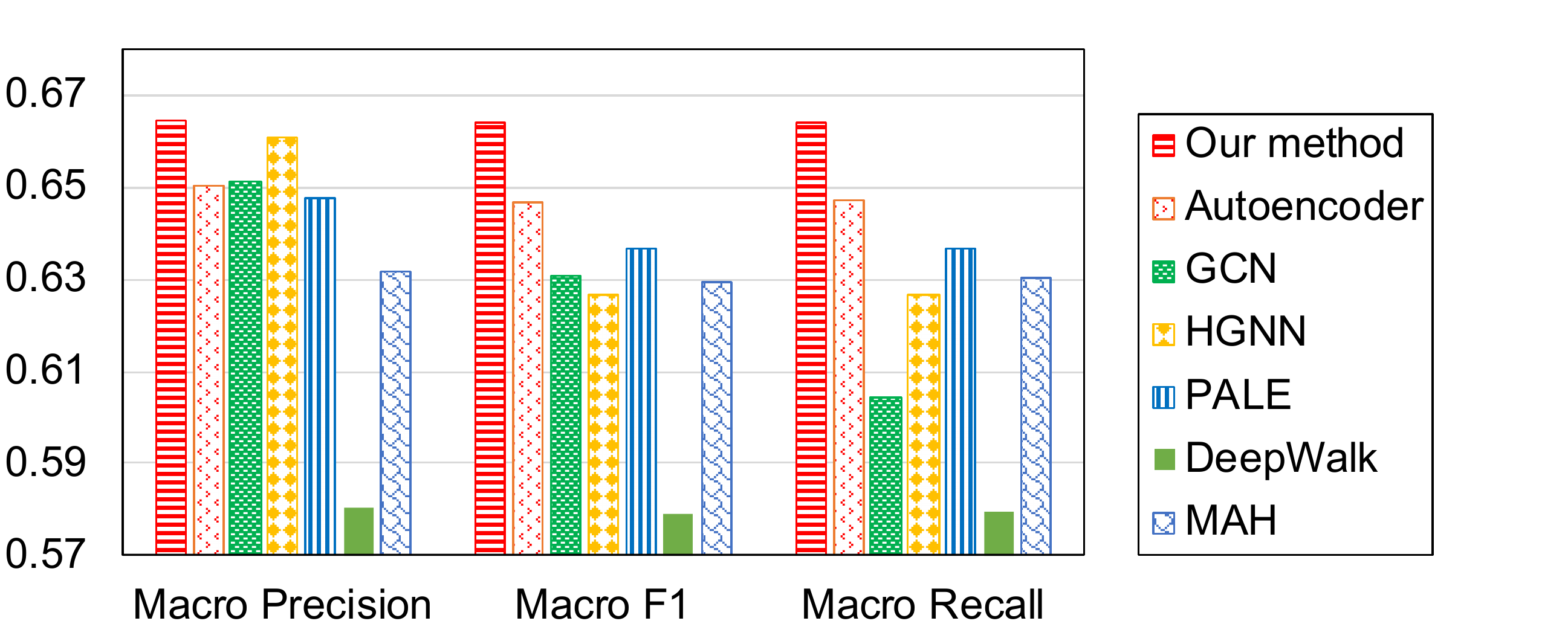}
}
\vspace{-1.4em}
\caption{Results on anchor link prediction.}
\label{fig:social_link}
\vspace{-2em}
\end{figure*}

\subsection{Datasets}
\label{sub:datasets}
\par For anchor link prediction, we use two cross-platform datasets collected and published in previous research on aligning heterogenous social networks \cite{cao2016bass}. One is the \textbf{Facebook-Twitter} dataset, and the other is the \textbf{Douban-Weibo} dataset. Facebook-Twitter contains 1,091,489 nodes, where the Facebook network has 422,291 nodes and 3,710,789 social links while the Twitter network contains 669,198 nodes that are connected by 12,749,257 social links. In Facebook-Twitter, 328,224 aligned user pairs are identified across two networks. Douban-Weibo bridges two popular social media platforms in China, namely Douban with 141,614 nodes and 2,700,602 social links and Weibo with 141,614 nodes with 6,280,561 social links. There are 141,614 aligned users in the total 283,228 nodes across these two networks in the Douban-Weibo dataset.

\par For parameter sensitivity and robustness analysis on anchor link prediction, we follow \cite{man2016predict} to generate two sub-networks from Facebook.
Specifically, we define a sparsity parameter $\alpha_s$ to control the sample ratio of edges from the original Facebook network, and $\alpha_c$ to control the ratio of shared edges in two sub-networks. For each edge, we generate a random value $p$ in $[0,1]$. If $p \leq 1-2\alpha_s+\alpha_s\alpha_c $, the edge is discarded; If $1-2\alpha_s+\alpha_s\alpha_c <p \leq 1-\alpha_s$, it is added in the first sub-network; If $1-\alpha_s<p \leq 1-\alpha_s\alpha_c$, it is only kept in the second sub-network; Otherwise, the edge is added in both sub-networks.
The reason of using extracted sub-networks instead of the full dataset is that we can customize the network sparsity via $\alpha_s$, and the node overlap level via $\alpha_c$. Hence, the flexible compositions of generated datasets can simulate a wide range of different application scenarios for testing different models' performance. Besides, they are relatively smaller than Facebook-Twitter and Douban-Weibo, thus enabling running time reduction for parameter sensitivity analysis.
\vspace{-1em}
\subsection{Baseline Methods}\vspace{-.5em}
We compare our method against the following baselines:

\begin{itemize}
	\item \textbf{Autoencoder} \cite{salha2019gravity}. This method uses one-hot encodings of nodes as the input and learns node representations by optimizing the mean square error loss function.
	\item \textbf{MAH} \cite{tan2014mapping}. This method enforces that a pair of nodes in the same hyperedge should come closer to learn node representations for anchor link prediction.
	\item \textbf{DeepWalk} \cite{perozzi2014deepwalk}. This method uses random walk to sample node sequences, and then learns node embeddings with the word2vec model. 
	\item \textbf{GCN} \cite{defferrard2016convolutional}. This method defines convolutional networks on graphs for node representation learning.
	\item \textbf{PALE} \cite{man2016predict}. This method predicts anchor links via network embedding by maximizing the log likelihood of observed edges and latent space matching. 
	\item \textbf{HGNN} \cite{feng2019hypergraph}. This method proposes hypergraph convolutional networks for network embedding. 
\end{itemize}

It is worth mentioning that the baselines we have chosen are all network embedding-based. In both datasets, the user profile and content information are unavailable, making traditional methods \cite{iofciu2011identifying, malhotra2012studying, liu2013s} that rely on auxiliary data sources inapplicable.

\vspace{-1.2em}
\subsection{Experimental Settings}
\subsubsection{\textbf{Evaluation Metrics}}
\par Following related works \cite{man2016predict,liu2013s}, we treat anchor link prediction as a binary classification task. Specifically, with a pair of nodes $(u,v)$ as input, we aim to predict whether they represent the same entity in two networks or not. As such, we leverage three widely-used classification metrics, namely Macro Precision, Macro Recall, and Macro F1. 
%
%
\vspace{-.8em}
\subsubsection{\textbf{Parameter Settings}}
\par For anchor link prediction, the ratio of positive and negative anchor links is set to $1:1$ for both the training and test. We train all methods using 50\% of the positive and negative links and test them on the remaining portion. In the graph partition and reconciliation step, we set $N_{min} = 1,000$, $N_{max} = 15,000$, and $N = 1,000$. The layer size $K$ is $2$ in our model. We construct the hypergraph via each node's 10 hop neighbors. That is, we connect each node and its 10 hop neighbors with one hyperedge. Note that we also adopt three other hypergraph construction strategies, and their impact will be discussed in section \ref{subsec:hyperconst}. The learning rate and embedding dimension are respectively fixed to 0.01 and 200 in our model. The negative link number in Equation~(\ref{equ:emb_loss}) is set to $M=5$. For all baseline methods, we adopt their reported optimal parameters by default.

\begin{table*}[h!]
\vspace{-1em}
\caption{Experimental results under different sparsity levels.}
\vspace{-1em}
\label{tab:result_as}
\resizebox{0.9\textwidth}{!}{%
\renewcommand{\arraystretch}{0.9}
\setlength\tabcolsep{10pt}
\begin{tabular}{@{}ll|ccccccccc@{}}
\toprule
\multicolumn{1}{l}{}                              &             & \multicolumn{9}{c}{sparsity level $\alpha_s$}                                                                                                                              \\ \midrule
Metric                                            & Model       & 10\%             & 20\%             & 30\%             & 40\%             & 50\%             & 60\%             & 70\%             & 80\%             & 90\%             \\ \midrule
\multirow{5}{*}{Macro Precision}                  & Our method  & \textbf{0.8620}  & \textbf{0.9071} & \textbf{0.9353} & \textbf{0.9345} & \textbf{0.9440}  & \textbf{0.9631} & \textbf{0.9638} & \textbf{0.9624} & \textbf{0.9638} \\
                                                  & Autoencoder & 0.8338          & 0.8455          & 0.8601          & 0.8195          & 0.8336          & 0.8590           & 0.9204          & 0.9255          & 0.8819          \\
                                                  & GCN         & 0.8340           & 0.8457          & 0.8881          & 0.8862          & 0.9115          & 0.9252          & 0.9366          & 0.9359          & 0.9434          \\
                                                  & HGNN        & 0.7295          & 0.8334          & 0.8340           & 0.8351          & 0.8376          & 0.8770           & 0.9025          & 0.8787          & 0.8850           \\
                                                  & PALE        & 0.8334          & 0.8337          & 0.7333          & 0.8337          & 0.8337          & 0.8338          & 0.8336          & 0.7648          & 0.7711          \\ \midrule
\multirow{5}{*}{Macro F1}                         & Our method  & \textbf{0.8602} & \textbf{0.9110}  & \textbf{0.9418} & \textbf{0.9438} & \textbf{0.9523} & \textbf{0.9701} & \textbf{0.9705} & \textbf{0.9698} & \textbf{0.9713} \\
                                                  & Autoencoder & 0.7450           & 0.8499          & 0.8603          & 0.8273          & 0.8377          & 0.8685          & 0.9247          & 0.9337          & 0.8924          \\
                                                  & GCN         & 0.7351          & 0.8030           & 0.8583          & 0.8721          & 0.9101          & 0.9250           & 0.9347          & 0.9386          & 0.9406          \\
                                                  & HGNN        & 0.6667          & 0.7634          & 0.8064          & 0.8394          & 0.8459          & 0.8849          & 0.9123          & 0.8881          & 0.8954          \\
                                                  & PALE        & 0.6584          & 0.7078          & 0.7141          & 0.7327          & 0.7496          & 0.7534          & 0.7457          & 0.7581          & 0.7512          \\ \midrule
\multirow{5}{*}{Macro Recall}  & Our method  & \textbf{0.8615} & \textbf{0.9158} & \textbf{0.9512} & \textbf{0.9570}  & \textbf{0.9660}  & \textbf{0.9788} & \textbf{0.9788} & \textbf{0.9790}  & \textbf{0.9805} \\
\multicolumn{1}{l}{}                              & Autoencoder & 0.7608          & 0.8635          & 0.8760           & 0.8562          & 0.8715          & 0.8955          & 0.9337          & 0.9477          & 0.9165          \\
\multicolumn{1}{l}{}                              & GCN         & 0.7190           & 0.7897          & 0.8448          & 0.8678          & 0.9087          & 0.9247          & 0.9345          & 0.9423          & 0.9393          \\
\multicolumn{1}{l}{}                              & HGNN        & 0.6600            & 0.7705          & 0.8225          & 0.8570           & 0.8633          & 0.9005          & 0.9292          & 0.9067          & 0.9153          \\
\multicolumn{1}{l}{}                              & PALE        & 0.6502          & 0.6993          & 0.7065          & 0.7283          & 0.7430           & 0.7470           & 0.7372          & 0.7550           & 0.7417          \\ \bottomrule
\end{tabular}%
}
\begin{tablenotes}
        \footnotesize
        \item[] Entries in \textbf{bold} are the best results. For the sparsity level, a lower $\alpha_s$ leads to a sparser dataset. $\alpha_c$ is fixed to 0.6 in this test.
\end{tablenotes}

\vspace{-1em}  
\end{table*}
\begin{figure*}[h!]
\vspace{-1em}
\centering
\subfloat[Macro Precision]{
\label{pre_at}
\includegraphics[width=0.31\textwidth]{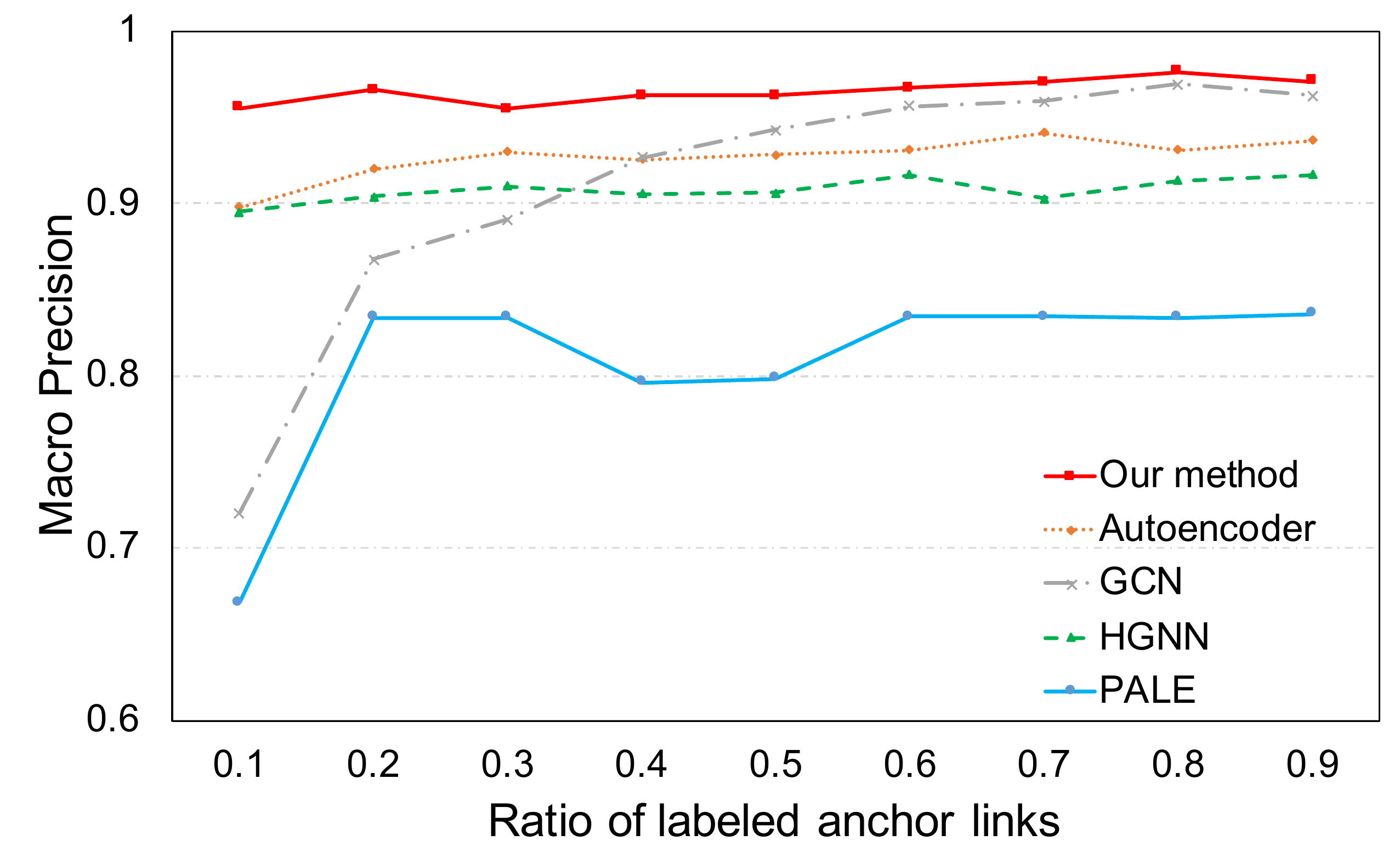}%
}
\subfloat[Macro F1]{
\label{ext}
\includegraphics[width=0.31\textwidth]{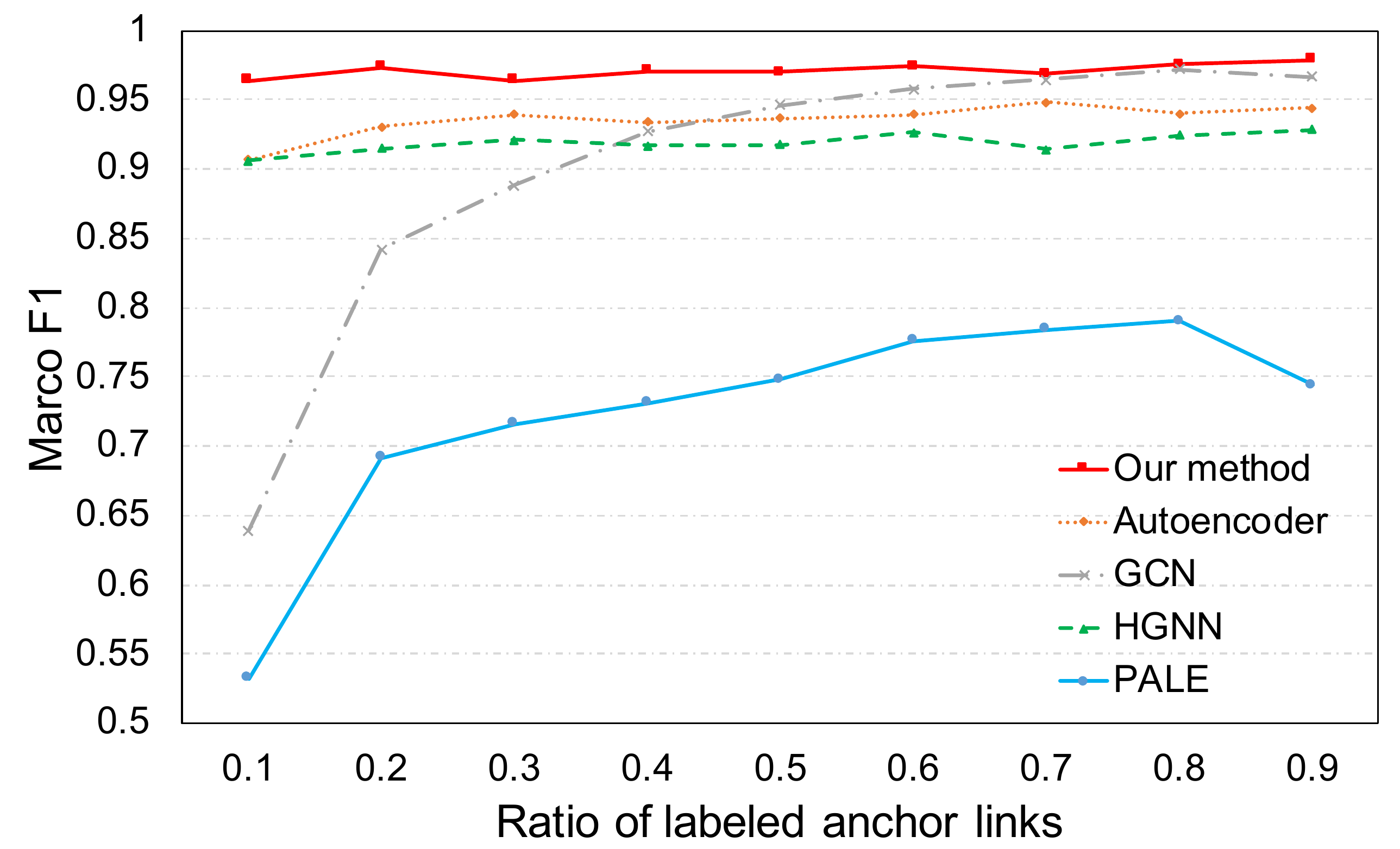}%
}
\subfloat[Macro Recall]{
\label{ext}
\includegraphics[width=0.31\textwidth]{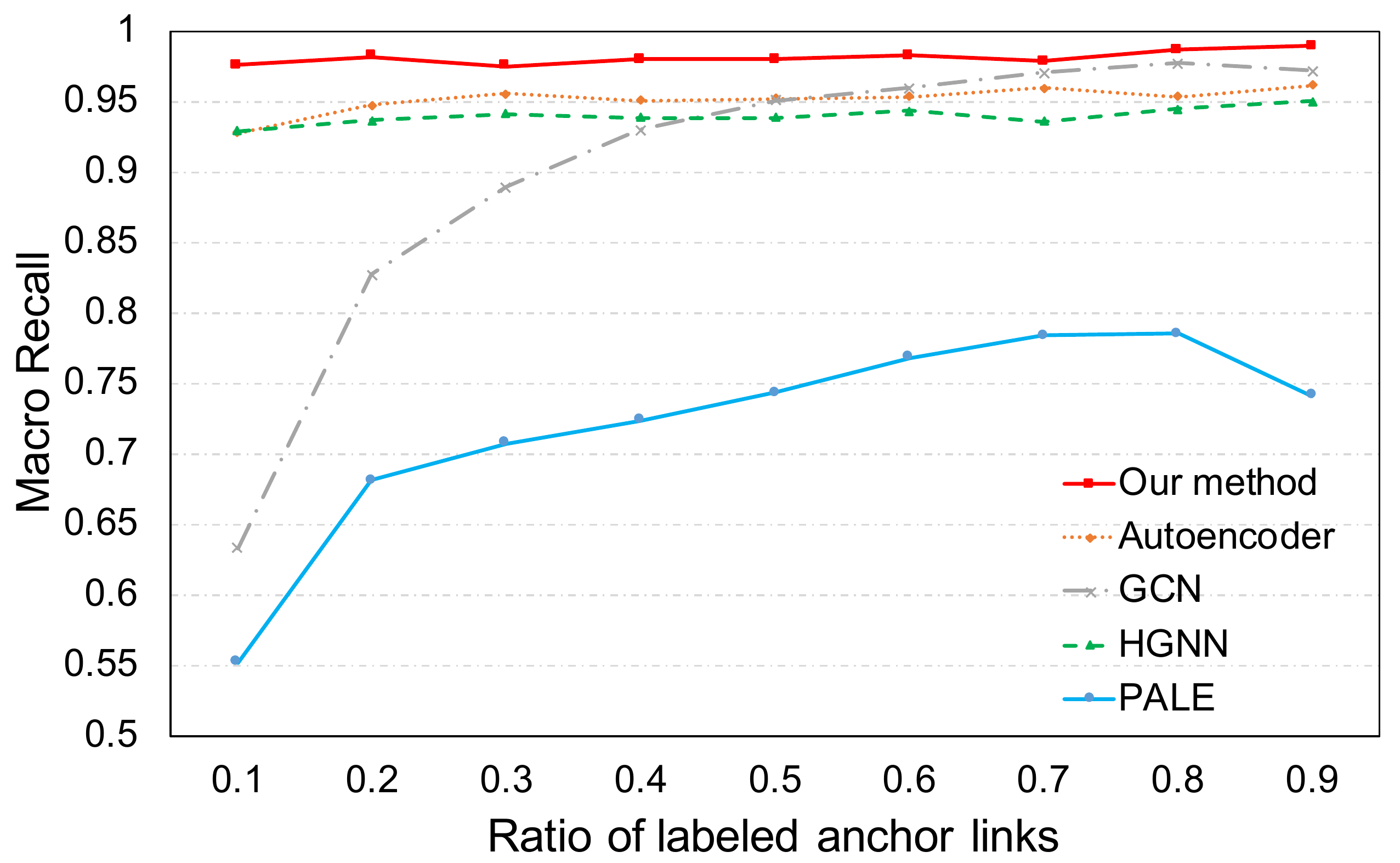}%
}\\
\vspace{-1.5em}
\caption{Results w.r.t. observed anchor link percentage.}
\label{fig:a_t}
\vspace{-1.5em}
\end{figure*}

\vspace{-1em}
\subsection{Performance on Anchor Link Prediction}
\label{sub:so_li}
\par In this section, we evaluate all models' performance on anchor link prediction on Facebook-Twitter and Douban-Weibo datasets. We report Macro Precision, F1, and Recall in Figure \ref{fig:social_link}. We draw the following observations.

Firstly, in terms of all evaluation metrics, our method has consistently and significantly outperformed all baselines on both datasets. Specifically, compared with the second best results on Macro Precision, Macro F1 and Macro Recall, our proposed MGCN achieves an improvement of 9.7\%, 9.1\%, and 9.0\% on Facebook-Twitter, and 0.6\%, 2.7\%, 2.6\% on Douban-Weibo, respectively. On one hand, MGCN performs both local graph convolution and hypergraph convolution operations on social networks, so it can effectively preserve the structural information in the learned node embeddings, leading to superior classification performance. On the other hand, traditional network embedding-based methods (e.g., DeepWalk and PALE) are unable to capture the complex, high-order node relationships, and tend to underperform on large-scale networks.

Secondly, as hypergraph-based baseline methods, HGNN shows stronger performance than MAH on both datasets. This is because HGNN largely benefits from the nonlinearity of neural networks, which offers higher model expressiveness while modeling hyperedges. Compared with GCN and PALE that only consider pairwise relations, both our method and HGNN can achieve better performance regarding Macro Precision and Macro Recall. This observation indicates the advantages of exploring hypergraphs for anchor link prediction. However, compared with both baselines, MGCN further incorporates node information extracted from local neighbourhood, thus enriching the granularity of learned node embeddings and yielding more competitive results. 

Thirdly, we also notice that our method is more advantageous on Facebook-Twitter than on Douban-Weibo. One possible reason is that the Douban-Weibo dataset has relatively higher density compared with Facebook-Twitter. When handling sparser datasets, GCN, DeepWalk and PALE suffer from severe performance decrease because they heavily rely on sufficient observed pairwise relations for node representation learning. This further demonstrates that our MGCN maintains high-level performance and shows promising robustness in the presence of data sparsity problem.

\vspace{-1.1em}
\subsection{Analysis on Model Robustness}
\vspace{-0.2em}
As we have previously mentioned, existing anchor link prediction methods are prone to suffer from performance downgrade when exposed to sparse datasets, and our proposed MGCN alleviates this problem by thoroughly investigating structural information within both simple graphs and the extracted hypergraphs. To test the robustness of our model, we carry out further comparisons with baselines on the two subnetworks extracted from Facebook network. To be specific, we vary the data compositions in these two subnetworks by adjusting the proportions of training labels (i.e., observed anchor links), edges (i.e., user-user pairwise interactions) and network overlaps (i.e., shared same nodes), and record the performance fluctuations of different models. We choose Autoencoder, GCN, HGNN and PALE in this analysis as they have competitive overall effectiveness and are relatively stable on large-scale datasets.
\vspace{-.8em}
\subsubsection{\textbf{Effect of Anchor Link Percentage}} \label{subsub:observed_per} 
In practice, the availability of the observed anchor nodes between two social networks that can be used for training are usually very limited. To test the impact of available anchor links, we firstly hold out $10\%$ of the observed anchor links for test, and change the ratio of anchor links from $10\%$ to $90\%$ for training. Note that two parameters $\alpha_s$ and $\alpha_c$ are both fixed to 0.9 during this test. All experiments including the sampling are executed five times.  
We report the average results of our method and baselines in Figure \ref{fig:a_t}, from which we can see that even with a small portion of training labels, our method still performs the best compared with other baselines. This is particularly important because in the real-world, anchor links are often sparsely observed, thus our method is the most competitive choice when there are insufficient labels for training.
\vspace{-.5em}
\subsubsection{\textbf{Effect of Edge Percentage}} While most GCN-based methods heavily rely on the information passed along edges for node representation learning, most real-life networks are naturally sparse in terms of the number of edges. So, we evaluate our method and baselines by adjusting the sparsity parameter $\alpha_s$ mentioned in section \ref{sub:datasets} from 10\% to 90\%, and report the everage results achieved in five executions as well. Note that we still use the same evaluation set as in Section \ref{subsub:observed_per}. As shown in Table \ref{tab:result_as}, our method keeps stable w.r.t. different values of $\alpha_s$. The reason is that modeling hypergraphs on top of simple graphs with MGCN can provide additional structural information when the availability of edges in physical networks is limited. 
\vspace{-1em}
\subsubsection{\textbf{Effect of Network Overlap Percentage}} Network overlap refers to shared entities (users in our case) in two different networks, and the shared entities tend to have similar local neighborhood structures \cite{bayati2009algorithms} in both networks. It characterizes the homogeneity of two independent networks. In this section, we change the parameter $\alpha_c$ from 10\% to 90\% and show the average results of five executions in Table \ref{tab:overlap}, from which we notice even in 10\% overlap level, our method still keeps the best performance, and the superiority of our method becomes more obvious when $\alpha_c$ is larger.
\begin{table*}[h!]
\vspace{-1em}
\caption{Experimental results under different overlap levels.}
\vspace{-1em}
\label{tab:overlap}
\resizebox{0.9\textwidth}{!}
{%
\renewcommand{\arraystretch}{0.9}
\setlength\tabcolsep{10pt}
\begin{tabular}{@{}ll|lllllllll@{}}
\toprule
                                                     &                                  & \multicolumn{9}{c}{overlap level $\alpha_c$}                                                                                                                               \\ \midrule
Metric                                               & \multicolumn{1}{l|}{Model}       & 10\%            & 20\%            & 30\%            & 40\%            & 50\%            & 60\%            & 70\%            & 80\%            & 90\%            \\ \midrule
\multirow{5}{*}{Macro Precision}  & \multicolumn{1}{l|}{Our method}  & \textbf{0.8719} & \textbf{0.9176} & \textbf{0.9414} & \textbf{0.9495} & \textbf{0.9557} & \textbf{0.9613} & \textbf{0.9587} & \textbf{0.9641} & \textbf{0.9541} \\
\multicolumn{1}{c}{}                                 & \multicolumn{1}{l|}{Autoencoder} & 0.8334          & 0.8799          & 0.8501          & 0.9100          & 0.9112          & 0.8600          & 0.8683          & 0.8784          & 0.8969          \\
\multicolumn{1}{c}{}                                 & \multicolumn{1}{l|}{GCN}         & 0.8336          & 0.8726          & 0.9023          & 0.9016          & 0.9276          & 0.9318          & 0.9270          & 0.9427          & 0.9381          \\
\multicolumn{1}{c}{}                                 & \multicolumn{1}{l|}{HGNN}        & 0.8015          & 0.8343          & 0.8259          & 0.8615          & 0.8548          & 0.8848          & 0.8993          & 0.8850          & 0.8902          \\
\multicolumn{1}{c}{}                                 & \multicolumn{1}{l|}{PALE}        & 0.8340          & 0.8015          & 0.8334          & 0.8334          & 0.8337          & 0.8336          & 0.7623          & 0.8334          & 0.8334          \\ \midrule
\multirow{5}{*}{Macro F1}                            & \multicolumn{1}{l|}{Our method}  & \textbf{0.8779} & \textbf{0.9256} & \textbf{0.9499} & \textbf{0.9570} & \textbf{0.9640} & \textbf{0.9691} & \textbf{0.9670} & \textbf{0.9713} & \textbf{0.9630} \\
                                                     & \multicolumn{1}{l|}{Autoencoder} & 0.8250          & 0.8872          & 0.8592          & 0.9171          & 0.9134          & 0.8697          & 0.8781          & 0.8885          & 0.9074          \\
                                                     & \multicolumn{1}{l|}{GCN}         & 0.7795          & 0.8436          & 0.8864          & 0.8920          & 0.9228          & 0.9319          & 0.9282          & 0.9448          & 0.9378          \\
                                                     & \multicolumn{1}{l|}{HGNN}        & 0.6537          & 0.7968          & 0.8330          & 0.8706          & 0.8642          & 0.8954          & 0.9100          & 0.8942          & 0.9007          \\
                                                     & \multicolumn{1}{l|}{PALE}        & 0.6966          & 0.7407          & 0.7542          & 0.7467          & 0.7612          & 0.7616          & 0.7554          & 0.7668          & 0.7655          \\ \midrule
\multirow{5}{*}{Macro Recall}                        & \multicolumn{1}{l|}{Our method}  & \textbf{0.8870} & \textbf{0.9363} & \textbf{0.9617} & \textbf{0.9665} & \textbf{0.9748} & \textbf{0.9790} & \textbf{0.9775} & \textbf{0.9800} & \textbf{0.9748} \\
                                                     & \multicolumn{1}{l|}{Autoencoder} & 0.8560          & 0.8980          & 0.8822          & 0.9312          & 0.9272          & 0.8935          & 0.9005          & 0.9107          & 0.9260          \\
                                                     & \multicolumn{1}{l|}{GCN}         & 0.7762          & 0.8337          & 0.8787          & 0.8842          & 0.9213          & 0.9330          & 0.9295          & 0.9470          & 0.9380          \\
                                                     & \multicolumn{1}{l|}{HGNN}        & 0.6465          & 0.8088          & 0.8515          & 0.8880          & 0.8850          & 0.9167          & 0.9290          & 0.9113          & 0.9210          \\
                                                     & PALE                             & 0.6885          & 0.7325          & 0.7480          & 0.7400          & 0.7590          & 0.7558          & 0.7520          & 0.7615          & 0.7632          \\ \bottomrule
\end{tabular}%
}
\begin{tablenotes}
        \footnotesize
        \item[] Entries in \textbf{bold} are the best results. For the overlap level, a higher $\alpha_c$ leads to more overlaps in two networks. $\alpha_s$ is fixed to 0.6 in this test.
\end{tablenotes}
\vspace{-1em}
\end{table*}

\begin{figure}[t]
\centering
\includegraphics[width=6.5cm,height=3.6cm]{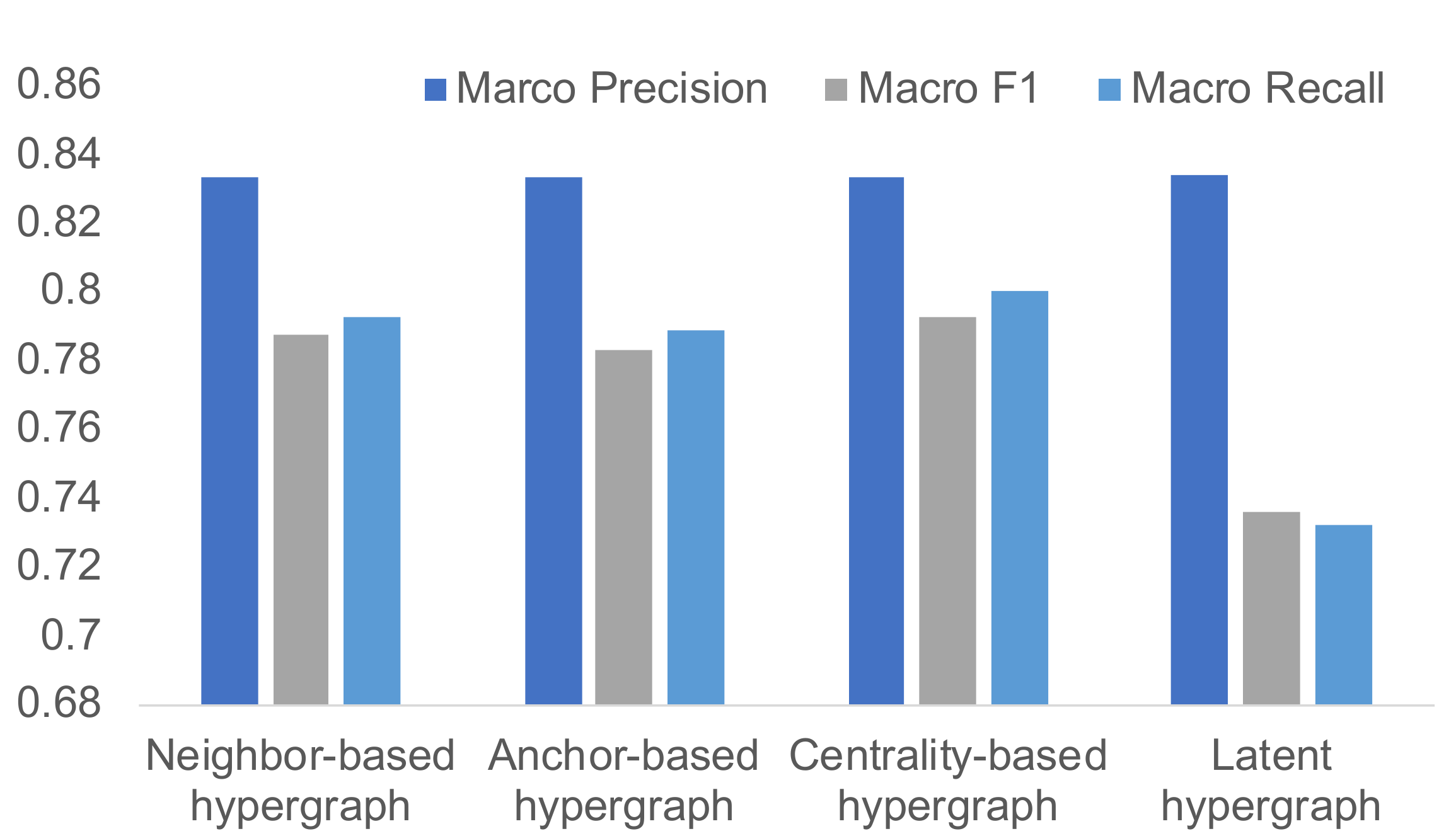}
\vspace{-1.2em}
\caption{Performance w.r.t. different hypergraph construction methods.}
\label{fig:hypertype}
\vspace{-3em}
\end{figure}
\vspace{-2.5em}
\subsection{Impact of Hypergraph Construction Strategies}\label{subsec:hyperconst}
We supply four methods for extracting hypergraphs from original networks, and compare their impacts to model performance below.
\begin{enumerate}
\vspace{-.5em}
	\item \textit{\textbf{Neighborhood-based hypergraph construction}}. This is the default hypergraph construction method we use for our experiment in Section \ref{sub:so_li}. For each node, we collect its $\phi$-hop neighbors and connect them in one hyperedge. As such, for a sub-graph with $N$ nodes, we finally have $N$ hyperedges. $\phi$ is optimized via grid search in $\{4,6,8,10,12\}$ and is set to 10 in our experiments.
	\item \textit{\textbf{Anchor-based hypergraph construction}}. This method is similar to the first one but we only consider the 10-hop neighbours of anchor nodes. That means, for a given sub-graph with $N$ nodes and $M$ observed anchor nodes, we will result in $M$ hyperedges. Since $M\ll N$ usually holds, this method is more practical when graph partitions are not applied on large-scale graphs.
	\item \textit{\textbf{Centrality-based hypergraph construction}}. We compute the following centrality values for each node: degree, betweenness, clustering coefficient, eigenvector, page rank, closeness centrality, node clique number, and communities a node belongs to. With these centrality-based properties, we generate a 20-dimensional vector (8-bit centrality-based features and 12-bit one-hot community encodings) for each node. By treating each dimension of the vector as a hyperedge, then each node's value on a specific dimension denotes the probability that this node belongs to the hyperedge.
	\item \textit{\textbf{Latent feature-based hypergraph construction}}. This Strategy uses Autoencoder to extract dense latent representations of nodes (we set the latent dimension to 200), where each latent dimension serves as a hyperedge. 
\end{enumerate}

The performance w.r.t. different hypergraph construction strategies are shown in Figure \ref{fig:hypertype}. We set $\alpha_c=\alpha_s=0.3$ for this test. In general, neighbor-based, anchor-based, and centrality-based hypergraphs lead to very close results. Surprisingly, though centrality-based hypergraph construction strategy involves carefully handcrafted features, it falls short in terms of Macro F1 and Macro Recall. This suggests that we do not have to design specific features to obtain performance improvements, which makes our method more practical for large datasets.
\vspace{-1em}

\begin{figure}[h]
\vspace{-.1em}
\centering
\includegraphics[width=6.2cm,height=4.3cm]{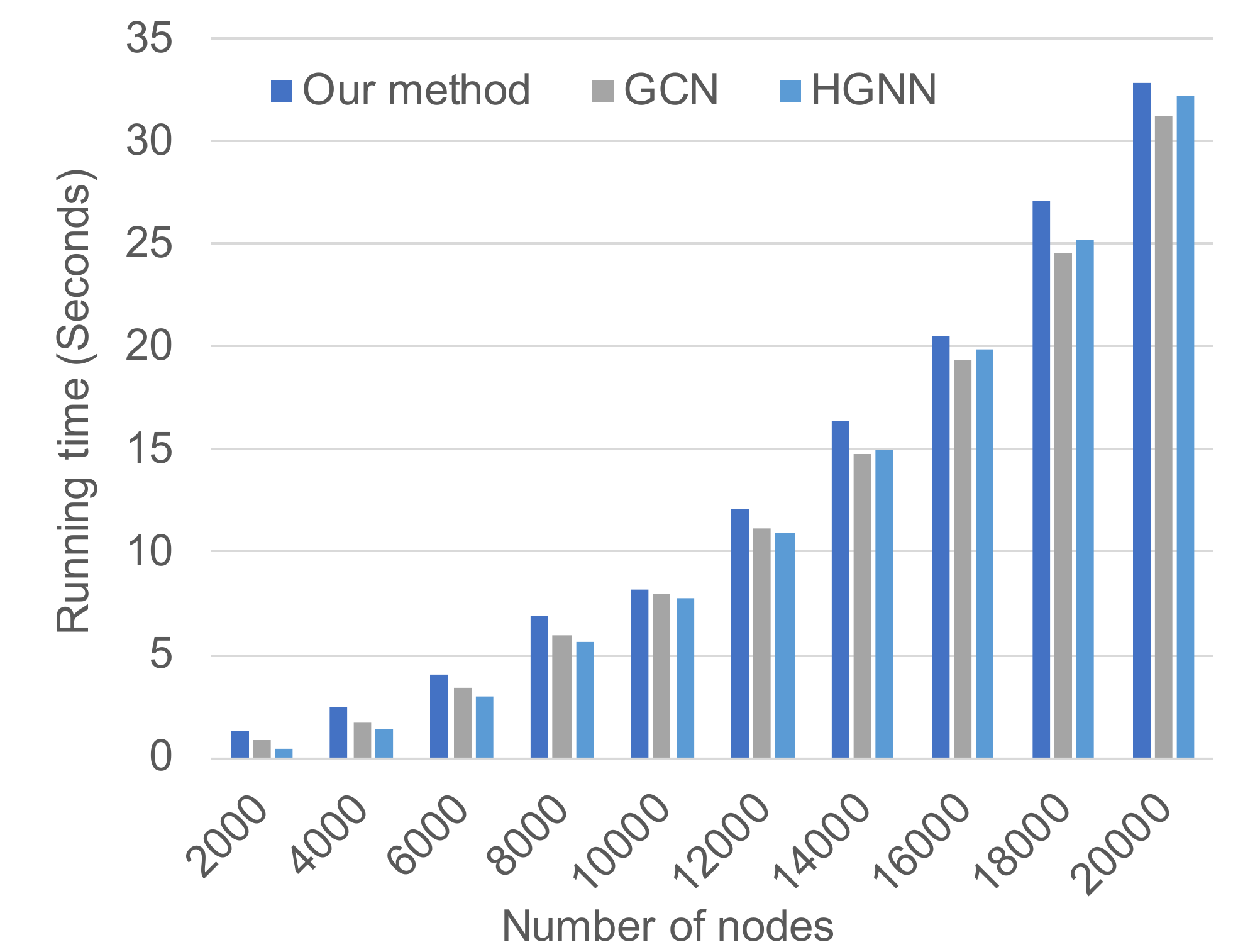}
\vspace{-1.2em}
\caption{Forward propagation time w.r.t. network scales.}
\label{fig:time}
\vspace{-2em}
\end{figure}
\vspace{-.15cm}
\subsection{Analysis on Model Efficiency}
To showcase the efficiency of MGCN, we calculate the running time of forward propagation for 1,000 epochs w.r.t. an increasing scale of the Facebook subnetworks and compare it with GCN and HGNN. All experiments are conducted on a linux server with two GTX Titan GPUs. The results are shown in Figure \ref{fig:time}. From the model architecture perspective, our method involves two GCN operations on both simple graphs and hypergraphs. However, compared with GCN and HGNN that only models simple graphs or hypergraphs, we can find that there is only a little additional time consumption of MGCN. This verifies the necessity and efficacy of modeling hyperedges in parallel. As a result, though MGCN achieves significant performance gain over all baselines, it still has very close efficiency to GCN and HGNN. Hence, for even larger datasets, our method can offer state-of-the-art anchor link prediction performance while retaining high-level scalability. 

\vspace{-2em}
\section{Related work}
\vspace{-.5em}
\subsection{Anchor Link Prediction in Social Networks}

\textbf{Traditional methods.} Traditionally, early studies solve the problem of account matching by leveraging user profile (e.g., user name, age, location) and their generated contents such as textual reviews and posts \cite{iofciu2011identifying, malhotra2012studying, liu2013s, goga2015reliability}. However, due to the difficulty of obtaining high-quality and credible data from the Internet, these methods inevitably suffer from the data insufficiency problem. As a result, these methods cannot achieve satisfactory results, and are subject to constrained generalizability in practice. Other techniques adopt matrix factorization to directly compute an alignment matrix \cite{trung2020comparative}, such as IsoRank \cite{singh2008global}, NetAligh \cite{bayati2009algorithms}, FINAL\cite{zhang2016final}, and REGAL\cite{heimann2018regal}. However, such approaches can hardly scale up to very large networks, because they take the entire adjacency matrices of networks as their input, which is highly demanding on storage and computing resources. Furthermore, they are prone to struggle when handling higher sparsity that comes with large-scale networks.\\
\textbf{Embedding-based approaches.} There have been applications of account matching by using network embedding techniques\cite{wang2019online,chen2018effective}. PALE \cite{man2016predict} learns node embedding by maximizing the co-occurrence likelihood of connected vertices, then applies linear projection or multi-layer perceptron (MLP) as the mapping function. Similar methods also include IONE \cite{liu2016aligning} which addresses this problem by modeling user-user following relationships in social networks. Though DALAUP \cite{cheng2019deep} further employs active learning to learn node embeddings, it is limited by its scalability as the active learning scheme can be time-consuming on large-scale social networks. DeepLink \cite{zhou2018deeplink} employs unbiased random walk to generate embeddings using skip-gram, then adopts auto-encoder and MLP as the mapping function. Manifold Alignment on Hypergraph (MAH) \cite{tan2014mapping} uses hypergraphs to model high-order relations by exploiting the idea that a pair of nodes in the same hyperedge should come closer. MAH is a pure hypergraph-based approach, which is simple and effective. However, it only considers sub-space learning for hyperedges, and is therefore vulnerable to noises and the loss of important underlying network structure information. In contrast, our proposed method proposes a decentralized hypergraph representation learning scheme, thus being able to handle large-scale social networks with a novel subgraph reconciliation mechanism. 
\vspace{-1em}
\subsection{Network Embedding}
Mainstream network embedding approaches include matrix factorization based methods, such as Multi Dimensional Scaling (MDS) \cite{mds}, Spectral Clustering \cite{spectral}, Graph Factorization \cite{ahmed2013distributed}, etc., as well as random walk-based methods \cite{tang2015line,perozzi2014deepwalk, grover2016node2vec} which firstly sample random walk node sequences, and then learn node embeddings via the skip-gram model. The recently proposed Graph Convolutional Networks (GCNs) \cite{hamilton2017inductive,chen2019exploiting} successfully define convolutional kernels on graph-structured data to learn node representations by aggregating information passed from its surrounding neighbours. More recently, different from traditional GCNs that only model simple graphs, hypergraphs have been infused into the context of graph convolutions \cite{yadati2019hypergcn, feng2019hypergraph}, enabling the learning of richer structural information. HGCN \cite{feng2019hypergraph} introduces the concept of hypergraph Laplacian, and then proposes a hypergraph-based extension to the original convolution on simple graphs. HyperGCN \cite{yadati2019hypergcn} also trains GCNs on hypergraphs with the utilization of hypergraph spectral theory. In this paper, we develop a specific GCN-like model that innovatively facilitates GCN operations at both hypergraph-level and simple graph-level in a unified framework to allow for comprehensive node representation learning.
\vspace{-0.5em}

\vspace{-0.5em}
\section{Conclusion}\label{sec:con}
We propose a multi-level graph convolution networks for anchor link prediction. Through the fusion of simple graph and hypergraph, our method steadily outperforms state-of-the-art methods. To handle large scale dataset, we also design a framework with network partitioning and two-phases reconciliation. 
The future work would suggest to explore the automatic discovery of hypergraphs for account matching problems, as well as scaling our framework to multiple social networks. Moreover, considering how to leverage the unused links between partitions are also potential ways of improving this work. Applications of this work with temporal analysis and recommendations \cite{chen2020sequence,chen2020try} will also be our future work.
\vspace{-1em}
\section*{Acknowledgment}
The work has been supported by Australian Research Council (Grant No. DP190101087, DP190101985, DP170103954 and FT200100825).
\vspace{-2em}

\bibliographystyle{abbrv}

\begin{thebibliography}{}

\end{thebibliography}


\begin{thebibliography}{10}

\bibitem{ahmad2010link}
M.~A. Ahmad, Z.~Borbora, J.~Srivastava, and N.~Contractor.
\newblock Link prediction across multiple social networks.
\newblock In {\em ICDMW}. IEEE, 2010.

\bibitem{ahmed2013distributed}
A.~Ahmed, N.~Shervashidze, S.~Narayanamurthy, V.~Josifovski, and A.~J. Smola.
\newblock Distributed large-scale natural graph factorization.
\newblock In {\em WWW13}.

\bibitem{bayati2009algorithms}
M.~Bayati, M.~Gerritsen, D.~F. Gleich, A.~Saberi, and Y.~Wang.
\newblock Algorithms for large, sparse network alignment problems.
\newblock In {\em ICDM}. IEEE, 2009.

\bibitem{blondel2008fast}
V.~D. Blondel, J.-L. Guillaume, R.~Lambiotte, and E.~Lefebvre.
\newblock Fast unfolding of communities in large networks.
\newblock {\em Journal of statistical mechanics: theory and experiment}, 2008.

\bibitem{cao2016bass}
X.~Cao and Y.~Yu.
\newblock Bass: A bootstrapping approach for aligning heterogenous social
  networks.
\newblock In {\em Joint European Conference on Machine Learning and Knowledge
  Discovery in Databases}. Springer, 2016.

\bibitem{chen2019exploiting}
H.~Chen, H.~Yin, T.~Chen, Q.~V.~H. Nguyen, W.-C. Peng, and X.~Li.
\newblock Exploiting centrality information with graph convolutions for network
  representation learning.
\newblock In {\em ICDE}. IEEE, 2019.

\bibitem{chen2018pme}
H.~Chen, H.~Yin, W.~Wang, H.~Wang, Q.~V.~H. Nguyen, and X.~Li.
\newblock Pme: projected metric embedding on heterogeneous networks for link
  prediction.
\newblock In {\em KDD}, 2018.

\bibitem{chen2020sequence}
T.~Chen, H.~Yin, Q.~V.~H. Nguyen, W.-C. Peng, X.~Li, and X.~Zhou.
\newblock Sequence-aware factorization machines for temporal predictive
  analytics.
\newblock In {\em ICDE}, 2020.

\bibitem{chen2020try}
T.~Chen, H.~Yin, G.~Ye, Z.~Huang, Y.~Wang, and M.~Wang.
\newblock Try this instead: Personalized and interpretable substitute
  recommendation.
\newblock {\em arXiv preprint}, 2020.

\bibitem{chen2018effective}
W.~Chen, H.~Yin, W.~Wang, L.~Zhao, and X.~Zhou.
\newblock Effective and efficient user account linkage across location based
  social networks.
\newblock In {\em ICDE}, pages 1085--1096. IEEE, 2018.

\bibitem{cheng2019deep}
A.~Cheng, C.~Zhou, H.~Yang, J.~Wu, L.~Li, J.~Tan, and L.~Guo.
\newblock Deep active learning for anchor user prediction.
\newblock {\em IJCAI}, 2019.

\bibitem{defferrard2016convolutional}
M.~Defferrard, X.~Bresson, and P.~Vandergheynst.
\newblock Convolutional neural networks on graphs with fast localized spectral
  filtering.
\newblock In {\em NIPS}, 2016.

\bibitem{faisal2014global}
F.~E. Faisal, H.~Zhao, and T.~Milenkovi{\'c}.
\newblock Global network alignment in the context of aging.
\newblock {\em Transactions on Computational Biology and Bioinformatics}, 2014.

\bibitem{feng2019hypergraph}
Y.~Feng, H.~You, Z.~Zhang, R.~Ji, and Y.~Gao.
\newblock Hypergraph neural networks.
\newblock In {\em Proceedings of the AAAI Conference on Artificial
  Intelligence}, 2019.

\bibitem{goga2015reliability}
O.~Goga, P.~Loiseau, R.~Sommer, R.~Teixeira, and K.~P. Gummadi.
\newblock On the reliability of profile matching across large online social
  networks.
\newblock In {\em KDD}, 2015.

\bibitem{grover2016node2vec}
A.~Grover and J.~Leskovec.
\newblock node2vec: Scalable feature learning for networks.
\newblock In {\em KDD}, 2016.

\bibitem{hamilton2017inductive}
W.~Hamilton, Z.~Ying, and J.~Leskovec.
\newblock Inductive representation learning on large graphs.
\newblock In {\em NIPS}, 2017.

\bibitem{heimann2018regal}
M.~Heimann, H.~Shen, T.~Safavi, and D.~Koutra.
\newblock Regal: Representation learning-based graph alignment.
\newblock In {\em CIKM}, 2018.

\bibitem{iofciu2011identifying}
T.~Iofciu, P.~Fankhauser, F.~Abel, and K.~Bischoff.
\newblock Identifying users across social tagging systems.
\newblock In {\em AAAI Conference on Weblogs and Social Media}, 2011.

\bibitem{jiang2019dynamic}
J.~Jiang, Y.~Wei, Y.~Feng, J.~Cao, and Y.~Gao.
\newblock Dynamic hypergraph neural networks.
\newblock In {\em IJCAI}, pages 2635--2641, 2019.

\bibitem{kipf2016semi}
T.~N. Kipf and M.~Welling.
\newblock Semi-supervised classification with graph convolutional networks.
\newblock {\em ICLR}, 2017.

\bibitem{liu2013s}
J.~Liu, F.~Zhang, X.~Song, Y.-I. Song, C.-Y. Lin, and H.-W. Hon.
\newblock What's in a name? an unsupervised approach to link users across
  communities.
\newblock In {\em WSDM}, 2013.

\bibitem{liu2016aligning}
L.~Liu, W.~K. Cheung, X.~Li, and L.~Liao.
\newblock Aligning users across social networks using network embedding.
\newblock In {\em Ijcai}, pages 1774--1780, 2016.

\bibitem{malhotra2012studying}
A.~Malhotra, L.~Totti, W.~Meira~Jr, P.~Kumaraguru, and V.~Almeida.
\newblock Studying user footprints in different online social networks.
\newblock In {\em ASONAM}. IEEE, 2012.

\bibitem{man2015context}
T.~Man, H.~Shen, J.~Huang, and X.~Cheng.
\newblock Context-adaptive matrix factorization for multi-context
  recommendation.
\newblock In {\em CIKM}, 2015.

\bibitem{man2016predict}
T.~Man, H.~Shen, S.~Liu, X.~Jin, and X.~Cheng.
\newblock Predict anchor links across social networks via an embedding
  approach.
\newblock In {\em IJCAI}, 2016.

\bibitem{musial2013social}
K.~Musia{\l} and P.~Kazienko.
\newblock Social networks on the internet.
\newblock {\em WWW}, 2013.

\bibitem{narayanan2009anonymizing}
A.~Narayanan and V.~Shmatikov.
\newblock De-anonymizing social networks.
\newblock In {\em 2009 30th IEEE symposium on security and privacy}. IEEE,
  2009.

\bibitem{spectral}
A.~Y. Ng, M.~I. Jordan, and Y.~Weiss.
\newblock On spectral clustering: Analysis and an algorithm.
\newblock In {\em NIPS}, 2002.

\bibitem{perozzi2014deepwalk}
B.~Perozzi, R.~Al-Rfou, and S.~Skiena.
\newblock Deepwalk: Online learning of social representations.
\newblock In {\em KDD}, 2014.

\bibitem{riederer2016linking}
C.~Riederer, Y.~Kim, A.~Chaintreau, N.~Korula, and S.~Lattanzi.
\newblock Linking users across domains with location data: Theory and
  validation.
\newblock In {\em WWW}, 2016.

\bibitem{salha2019gravity}
G.~Salha, S.~Limnios, R.~Hennequin, V.-A. Tran, and M.~Vazirgiannis.
\newblock Gravity-inspired graph autoencoders for directed link prediction.
\newblock In {\em CIKM}, 2019.

\bibitem{singh2008global}
R.~Singh, J.~Xu, and B.~Berger.
\newblock Global alignment of multiple protein interaction networks with
  application to functional orthology detection.
\newblock {\em Proceedings of the National Academy of Sciences}, 2008.

\bibitem{tan2014mapping}
S.~Tan, Z.~Guan, D.~Cai, X.~Qin, J.~Bu, and C.~Chen.
\newblock Mapping users across networks by manifold alignment on hypergraph.
\newblock In {\em AAAI}, 2014.

\bibitem{tang2012etrust}
J.~Tang, H.~Gao, H.~Liu, and A.~Das~Sarma.
\newblock etrust: Understanding trust evolution in an online world.
\newblock In {\em KDD}, 2012.

\bibitem{tang2015line}
J.~Tang, M.~Qu, M.~Wang, M.~Zhang, J.~Yan, and Q.~Mei.
\newblock Line: Large-scale information network embedding.
\newblock In {\em WWW}, 2015.

\bibitem{trung2020comparative}
H.~T. Trung, N.~T. Toan, T.~Van~Vinh, H.~T. Dat, D.~C. Thang, N.~Q.~V. Hung,
  and A.~Sattar.
\newblock A comparative study on network alignment techniques.
\newblock {\em Expert Systems with Applications}, 2020.

\bibitem{wang2019online}
W.~Wang, H.~Yin, X.~Du, W.~Hua, Y.~Li, and Q.~V.~H. Nguyen.
\newblock Online user representation learning across heterogeneous social
  networks.
\newblock In {\em SIGIR}, 2019.

\bibitem{yadati2019hypergcn}
N.~Yadati, M.~Nimishakavi, P.~Yadav, V.~Nitin, A.~Louis, and P.~Talukdar.
\newblock Hypergcn: A new method for training graph convolutional networks on
  hypergraphs.
\newblock In {\em NIPS}, 2019.

\bibitem{yin2019social}
H.~Yin, Q.~Wang, K.~Zheng, Z.~Li, J.~Yang, and X.~Zhou.
\newblock Social influence-based group representation learning for group
  recommendation.
\newblock In {\em ICDE}. IEEE, 2019.

\bibitem{yin2018joint}
H.~Yin, L.~Zou, Q.~V.~H. Nguyen, Z.~Huang, and X.~Zhou.
\newblock Joint event-partner recommendation in event-based social networks.
\newblock In {\em ICDE}. IEEE, 2018.

\bibitem{zhang2015integrated}
J.~Zhang and S.~Y. Philip.
\newblock Integrated anchor and social link predictions across social networks.
\newblock In {\em AAAI}, 2015.

\bibitem{zhang2016final}
S.~Zhang and H.~Tong.
\newblock Final: Fast attributed network alignment.
\newblock In {\em KDD}, 2016.

\bibitem{zhang2015cosnet}
Y.~Zhang, J.~Tang, Z.~Yang, J.~Pei, and P.~S. Yu.
\newblock Cosnet: Connecting heterogeneous social networks with local and
  global consistency.
\newblock In {\em KDD}, 2015.

\bibitem{zhou2018deeplink}
F.~Zhou, L.~Liu, K.~Zhang, G.~Trajcevski, J.~Wu, and T.~Zhong.
\newblock Deeplink: A deep learning approach for user identity linkage.
\newblock In {\em IEEE INFOCOM}. IEEE, 2018.

\bibitem{zhou2015cross}
X.~Zhou, X.~Liang, H.~Zhang, and Y.~Ma.
\newblock Cross-platform identification of anonymous identical users in
  multiple social media networks.
\newblock {\em TKDE}, 2015.

\bibitem{mds}
L.~Zlatkov.
\newblock Multidimensional scaling (mds).
\newblock 1978.

\end{thebibliography}

\end{document}